\documentclass[longauth]{aa}

\usepackage[varg]{txfonts}
\usepackage{graphicx}
\usepackage{epstopdf}
\usepackage{url}

\begin{document}
 
\title{HESS\,J1741$-$302: a hidden accelerator in the Galactic plane}

\makeatletter
\renewcommand*{\@fnsymbol}[1]{\ifcase#1\or*\or$\dagger$\or$\ddagger$\or**\or$\dagger\dagger$\or$\ddagger\ddagger$\fi}
\makeatother

\author{\footnotesize H.E.S.S. Collaboration
\and H.~Abdalla \inst{\ref{NWU}}
\and A.~Abramowski \inst{\ref{HH}}
\and F.~Aharonian \inst{\ref{MPIK},\ref{DIAS},\ref{NASRA}}
\and F.~Ait~Benkhali \inst{\ref{MPIK}}
\and E.O.~Ang\"uner\protect\footnotemark[1] \inst{\ref{IFJPAN}}
\and M.~Arakawa \inst{\ref{Rikkyo}}
\and C.~Armand \inst{\ref{LAPP}}
\and M.~Arrieta \inst{\ref{LUTH}}
\and M.~Backes \inst{\ref{UNAM}}
\and A.~Balzer \inst{\ref{GRAPPA}}
\and M.~Barnard \inst{\ref{NWU}}
\and Y.~Becherini \inst{\ref{Linnaeus}}
\and J.~Becker~Tjus \inst{\ref{RUB}}
\and D.~Berge \inst{\ref{DESY}}
\and S.~Bernhard \inst{\ref{LFUI}}
\and K.~Bernl\"ohr \inst{\ref{MPIK}}
\and R.~Blackwell \inst{\ref{Adelaide}}
\and M.~B\"ottcher \inst{\ref{NWU}}
\and C.~Boisson \inst{\ref{LUTH}}
\and J.~Bolmont \inst{\ref{LPNHE}}
\and S.~Bonnefoy \inst{\ref{DESY}}
\and P.~Bordas \inst{\ref{MPIK}}
\and J.~Bregeon \inst{\ref{LUPM}}
\and F.~Brun \inst{\ref{CENB}}
\and P.~Brun \inst{\ref{IRFU}}
\and M.~Bryan \inst{\ref{GRAPPA}}
\and M.~B\"{u}chele \inst{\ref{ECAP}}
\and T.~Bulik \inst{\ref{UWarsaw}}
\and M.~Capasso \inst{\ref{IAAT}}
\and S.~Caroff \inst{\ref{LLR}}
\and A.~Carosi \inst{\ref{LAPP}}
\and S.~Casanova\protect\footnotemark[1] \inst{\ref{IFJPAN},\ref{MPIK}}
\and M.~Cerruti \inst{\ref{LPNHE}}
\and N.~Chakraborty \inst{\ref{MPIK}}
\and R.C.G.~Chaves \inst{\ref{LUPM},\ref{CurieChaves}}
\and A.~Chen \inst{\ref{WITS}}
\and J.~Chevalier \inst{\ref{LAPP}}
\and S.~Colafrancesco \inst{\ref{WITS}}
\and B.~Condon \inst{\ref{CENB}}
\and J.~Conrad \inst{\ref{OKC},\ref{FellowConrad}}
\and I.D.~Davids \inst{\ref{UNAM}}
\and J.~Decock \inst{\ref{IRFU}}
\and C.~Deil \inst{\ref{MPIK}}
\and J.~Devin \inst{\ref{LUPM}}
\and P.~deWilt \inst{\ref{Adelaide}}
\and L.~Dirson \inst{\ref{HH}}
\and A.~Djannati-Ata\"i \inst{\ref{APC}}
\and A.~Donath \inst{\ref{MPIK}}
\and L.O'C.~Drury \inst{\ref{DIAS}}
\and J.~Dyks \inst{\ref{NCAC}}
\and T.~Edwards \inst{\ref{MPIK}}
\and K.~Egberts \inst{\ref{UP}}
\and G.~Emery \inst{\ref{LPNHE}}
\and J.-P.~Ernenwein \inst{\ref{CPPM}}
\and S.~Eschbach \inst{\ref{ECAP}}
\and C.~Farnier \inst{\ref{OKC},\ref{Linnaeus}}
\and S.~Fegan \inst{\ref{LLR}}
\and M.V.~Fernandes \inst{\ref{HH}}
\and A.~Fiasson \inst{\ref{LAPP}}
\and G.~Fontaine \inst{\ref{LLR}}
\and S.~Funk \inst{\ref{ECAP}}
\and M.~F\"u{\ss}ling \inst{\ref{DESY}}
\and S.~Gabici \inst{\ref{APC}}
\and Y.A.~Gallant \inst{\ref{LUPM}}
\and T.~Garrigoux \inst{\ref{NWU}}
\and F.~Gat{\'e} \inst{\ref{LAPP}}
\and G.~Giavitto \inst{\ref{DESY}}
\and D.~Glawion \inst{\ref{LSW}}
\and J.F.~Glicenstein \inst{\ref{IRFU}}
\and D.~Gottschall \inst{\ref{IAAT}}
\and M.-H.~Grondin \inst{\ref{CENB}}
\and J.~Hahn \inst{\ref{MPIK}}
\and M.~Haupt \inst{\ref{DESY}}
\and J.~Hawkes \inst{\ref{Adelaide}}
\and G.~Heinzelmann \inst{\ref{HH}}
\and G.~Henri \inst{\ref{Grenoble}}
\and G.~Hermann \inst{\ref{MPIK}}
\and J.A.~Hinton \inst{\ref{MPIK}}
\and W.~Hofmann \inst{\ref{MPIK}}
\and C.~Hoischen \inst{\ref{UP}}
\and T.~L.~Holch \inst{\ref{HUB}}
\and M.~Holler \inst{\ref{LFUI}}
\and D.~Horns \inst{\ref{HH}}
\and A.~Ivascenko \inst{\ref{NWU}}
\and H.~Iwasaki \inst{\ref{Rikkyo}}
\and A.~Jacholkowska \inst{\ref{LPNHE}}
\and M.~Jamrozy \inst{\ref{UJK}}
\and D.~Jankowsky \inst{\ref{ECAP}}
\and F.~Jankowsky \inst{\ref{LSW}}
\and M.~Jingo \inst{\ref{WITS}}
\and L.~Jouvin \inst{\ref{APC}}
\and I.~Jung-Richardt \inst{\ref{ECAP}}
\and M.A.~Kastendieck \inst{\ref{HH}}
\and K.~Katarzy{\'n}ski \inst{\ref{NCUT}}
\and M.~Katsuragawa \inst{\ref{JAXA}}
\and U.~Katz \inst{\ref{ECAP}}
\and D.~Kerszberg \inst{\ref{LPNHE}}
\and D.~Khangulyan \inst{\ref{Rikkyo}}
\and B.~Kh\'elifi \inst{\ref{APC}}
\and J.~King \inst{\ref{MPIK}}
\and S.~Klepser \inst{\ref{DESY}}
\and D.~Klochkov \inst{\ref{IAAT}}
\and W.~Klu\'{z}niak \inst{\ref{NCAC}}
\and Nu.~Komin \inst{\ref{WITS}}
\and K.~Kosack \inst{\ref{IRFU}}
\and S.~Krakau \inst{\ref{RUB}}
\and M.~Kraus \inst{\ref{ECAP}}
\and P.P.~Kr\"uger \inst{\ref{NWU}}
\and H.~Laffon \inst{\ref{CENB}}
\and G.~Lamanna \inst{\ref{LAPP}}
\and J.~Lau \inst{\ref{Adelaide}}
\and J.~Lefaucheur \inst{\ref{LUTH}}
\and A.~Lemi\`ere \inst{\ref{APC}}
\and M.~Lemoine-Goumard \inst{\ref{CENB}}
\and J.-P.~Lenain \inst{\ref{LPNHE}}
\and E.~Leser \inst{\ref{UP}}
\and T.~Lohse \inst{\ref{HUB}}
\and M.~Lorentz \inst{\ref{IRFU}}
\and R.~Liu \inst{\ref{MPIK}}
\and R.~L\'opez-Coto \inst{\ref{MPIK}}
\and I.~Lypova \inst{\ref{DESY}}
\and D.~Malyshev \inst{\ref{IAAT}}
\and V.~Marandon \inst{\ref{MPIK}}
\and A.~Marcowith \inst{\ref{LUPM}}
\and C.~Mariaud \inst{\ref{LLR}}
\and R.~Marx \inst{\ref{MPIK}}
\and G.~Maurin \inst{\ref{LAPP}}
\and N.~Maxted \inst{\ref{Adelaide},\ref{MaxtedNowAt}}
\and M.~Mayer \inst{\ref{HUB}}
\and P.J.~Meintjes \inst{\ref{UFS}}
\and M.~Meyer \inst{\ref{OKC},\ref{MeyerNowAt}}
\and A.M.W.~Mitchell \inst{\ref{MPIK}}
\and R.~Moderski \inst{\ref{NCAC}}
\and M.~Mohamed \inst{\ref{LSW}}
\and L.~Mohrmann \inst{\ref{ECAP}}
\and K.~Mor{\aa} \inst{\ref{OKC}}
\and E.~Moulin \inst{\ref{IRFU}}
\and T.~Murach \inst{\ref{DESY}}
\and S.~Nakashima  \inst{\ref{JAXA}}
\and M.~de~Naurois \inst{\ref{LLR}}
\and H.~Ndiyavala  \inst{\ref{NWU}}
\and F.~Niederwanger \inst{\ref{LFUI}}
\and J.~Niemiec \inst{\ref{IFJPAN}}
\and L.~Oakes \inst{\ref{HUB}}
\and P.~O'Brien \inst{\ref{Leicester}}
\and H.~Odaka \inst{\ref{JAXA}}
\and S.~Ohm \inst{\ref{DESY}}
\and M.~Ostrowski \inst{\ref{UJK}}
\and I.~Oya \inst{\ref{DESY}}
\and M.~Padovani \inst{\ref{LUPM}}
\and M.~Panter \inst{\ref{MPIK}}
\and R.D.~Parsons \inst{\ref{MPIK}}
\and N.W.~Pekeur \inst{\ref{NWU}}
\and G.~Pelletier \inst{\ref{Grenoble}}
\and C.~Perennes \inst{\ref{LPNHE}}
\and P.-O.~Petrucci \inst{\ref{Grenoble}}
\and B.~Peyaud \inst{\ref{IRFU}}
\and Q.~Piel \inst{\ref{LAPP}}
\and S.~Pita \inst{\ref{APC}}
\and V.~Poireau \inst{\ref{LAPP}}
\and D.A.~Prokhorov \inst{\ref{WITS}}
\and H.~Prokoph \inst{\ref{GRAPPAHE}}
\and G.~P\"uhlhofer \inst{\ref{IAAT}}
\and M.~Punch \inst{\ref{APC},\ref{Linnaeus}}
\and A.~Quirrenbach \inst{\ref{LSW}}
\and S.~Raab \inst{\ref{ECAP}}
\and R.~Rauth \inst{\ref{LFUI}}
\and A.~Reimer \inst{\ref{LFUI}}
\and O.~Reimer \inst{\ref{LFUI}}
\and M.~Renaud \inst{\ref{LUPM}}
\and R.~de~los~Reyes \inst{\ref{MPIK}}
\and F.~Rieger \inst{\ref{MPIK},\ref{FellowRieger}}
\and L.~Rinchiuso \inst{\ref{IRFU}}
\and C.~Romoli \inst{\ref{DIAS}}
\and G.~Rowell \inst{\ref{Adelaide}}
\and B.~Rudak \inst{\ref{NCAC}}
\and C.B.~Rulten \inst{\ref{LUTH}}
\and V.~Sahakian \inst{\ref{YPI},\ref{NASRA}}
\and S.~Saito \inst{\ref{Rikkyo}}
\and D.A.~Sanchez \inst{\ref{LAPP}}
\and A.~Santangelo \inst{\ref{IAAT}}
\and M.~Sasaki \inst{\ref{ECAP}}
\and R.~Schlickeiser \inst{\ref{RUB}}
\and F.~Sch\"ussler \inst{\ref{IRFU}}
\and A.~Schulz \inst{\ref{DESY}}
\and U.~Schwanke \inst{\ref{HUB}}
\and S.~Schwemmer \inst{\ref{LSW}}
\and M.~Seglar-Arroyo \inst{\ref{IRFU}}
\and A.S.~Seyffert \inst{\ref{NWU}}
\and N.~Shafi \inst{\ref{WITS}}
\and I.~Shilon \inst{\ref{ECAP}}
\and K.~Shiningayamwe \inst{\ref{UNAM}}
\and R.~Simoni \inst{\ref{GRAPPA}}
\and H.~Sol \inst{\ref{LUTH}}
\and F.~Spanier \inst{\ref{NWU}}
\and M.~Spir-Jacob \inst{\ref{APC}}
\and {\L.}~Stawarz \inst{\ref{UJK}}
\and R.~Steenkamp \inst{\ref{UNAM}}
\and C.~Stegmann \inst{\ref{UP},\ref{DESY}}
\and C.~Steppa \inst{\ref{UP}}
\and I.~Sushch \inst{\ref{NWU}}
\and T.~Takahashi  \inst{\ref{JAXA}}
\and J.-P.~Tavernet \inst{\ref{LPNHE}}
\and T.~Tavernier \inst{\ref{IRFU}}
\and A.M.~Taylor \inst{\ref{DESY}}
\and R.~Terrier \inst{\ref{APC}}
\and L.~Tibaldo \inst{\ref{MPIK}}
\and D.~Tiziani \inst{\ref{ECAP}}
\and M.~Tluczykont \inst{\ref{HH}}
\and C.~Trichard \inst{\ref{CPPM}}
\and M.~Tsirou \inst{\ref{LUPM}}
\and N.~Tsuji \inst{\ref{Rikkyo}}
\and R.~Tuffs \inst{\ref{MPIK}}
\and Y.~Uchiyama \inst{\ref{Rikkyo}}
\and D.J.~van~der~Walt\protect\footnotemark[1] \inst{\ref{NWU}}
\and C.~van~Eldik \inst{\ref{ECAP}}
\and C.~van~Rensburg \inst{\ref{NWU}}
\and B.~van~Soelen \inst{\ref{UFS}}
\and G.~Vasileiadis \inst{\ref{LUPM}}
\and J.~Veh \inst{\ref{ECAP}}
\and C.~Venter \inst{\ref{NWU}}
\and A.~Viana \inst{\ref{MPIK},\ref{VianaNowAt}}
\and P.~Vincent \inst{\ref{LPNHE}}
\and J.~Vink \inst{\ref{GRAPPA}}
\and F.~Voisin \inst{\ref{Adelaide}}
\and H.J.~V\"olk \inst{\ref{MPIK}}
\and T.~Vuillaume \inst{\ref{LAPP}}
\and Z.~Wadiasingh \inst{\ref{NWU}}
\and S.J.~Wagner \inst{\ref{LSW}}
\and P.~Wagner \inst{\ref{HUB}}
\and R.M.~Wagner \inst{\ref{OKC}}
\and R.~White \inst{\ref{MPIK}}
\and A.~Wierzcholska \inst{\ref{IFJPAN}}
\and P.~Willmann \inst{\ref{ECAP}}
\and A.~W\"ornlein \inst{\ref{ECAP}}
\and D.~Wouters \inst{\ref{IRFU}}
\and R.~Yang \inst{\ref{MPIK}}
\and D.~Zaborov \inst{\ref{LLR}}
\and M.~Zacharias \inst{\ref{NWU}}
\and R.~Zanin \inst{\ref{MPIK}}
\and A.A.~Zdziarski \inst{\ref{NCAC}}
\and A.~Zech \inst{\ref{LUTH}}
\and F.~Zefi \inst{\ref{LLR}}
\and A.~Ziegler \inst{\ref{ECAP}}
\and J.~Zorn \inst{\ref{MPIK}}
\and N.~\.Zywucka \inst{\ref{UJK}}\\
NANTEN Collaboration
\and R.~Enokiya \inst{\ref{Nagoya}}
\and Y.~Fukui \inst{\ref{Nagoya}}
\and T.~Hayakawa \inst{\ref{Nagoya}} 
\and T.~Okuda \inst{\ref{Nobeya}}
\and K.~Torii \inst{\ref{Nagoya}}
\and H.~Yamamoto \inst{\ref{Nagoya}}
}

\institute{
Centre for Space Research, North-West University, Potchefstroom 2520, South Africa \label{NWU} \and 
Universit\"at Hamburg, Institut f\"ur Experimentalphysik, Luruper Chaussee 149, D 22761 Hamburg, Germany \label{HH} \and 
Max-Planck-Institut f\"ur Kernphysik, P.O. Box 103980, D 69029 Heidelberg, Germany \label{MPIK} \and 
Dublin Institute for Advanced Studies, 31 Fitzwilliam Place, Dublin 2, Ireland \label{DIAS} \and 
National Academy of Sciences of the Republic of Armenia,  Marshall Baghramian Avenue, 24, 0019 Yerevan, Republic of Armenia  \label{NASRA} \and
Yerevan Physics Institute, 2 Alikhanian Brothers St., 375036 Yerevan, Armenia \label{YPI} \and
Institut f\"ur Physik, Humboldt-Universit\"at zu Berlin, Newtonstr. 15, D 12489 Berlin, Germany \label{HUB} \and
University of Namibia, Department of Physics, Private Bag 13301, Windhoek, Namibia \label{UNAM} \and
GRAPPA, Anton Pannekoek Institute for Astronomy, University of Amsterdam,  Science Park 904, 1098 XH Amsterdam, The Netherlands \label{GRAPPA} \and
Department of Physics and Electrical Engineering, Linnaeus University,  351 95 V\"axj\"o, Sweden \label{Linnaeus} \and
Institut f\"ur Theoretische Physik, Lehrstuhl IV: Weltraum und Astrophysik, Ruhr-Universit\"at Bochum, D 44780 Bochum, Germany \label{RUB} \and
GRAPPA, Anton Pannekoek Institute for Astronomy and Institute of High-Energy Physics, University of Amsterdam,  Science Park 904, 1098 XH Amsterdam, The Netherlands \label{GRAPPAHE} \and
Institut f\"ur Astro- und Teilchenphysik, Leopold-Franzens-Universit\"at Innsbruck, A-6020 Innsbruck, Austria \label{LFUI} \and
School of Physical Sciences, University of Adelaide, Adelaide 5005, Australia \label{Adelaide} \and
LUTH, Observatoire de Paris, PSL Research University, CNRS, Universit\'e Paris Diderot, 5 Place Jules Janssen, 92190 Meudon, France \label{LUTH} \and
Sorbonne Universit\'es, UPMC Universit\'e Paris 06, Universit\'e Paris Diderot, Sorbonne Paris Cit\'e, CNRS, Laboratoire de Physique Nucl\'eaire et de Hautes Energies (LPNHE), 4 place Jussieu, F-75252, Paris Cedex 5, France \label{LPNHE} \and
Laboratoire Univers et Particules de Montpellier, Universit\'e Montpellier, CNRS/IN2P3,  CC 72, Place Eug\`ene Bataillon, F-34095 Montpellier Cedex 5, France \label{LUPM} \and
IRFU, CEA, Universit\'e Paris-Saclay, F-91191 Gif-sur-Yvette, France \label{IRFU} \and
Astronomical Observatory, The University of Warsaw, Al. Ujazdowskie 4, 00-478 Warsaw, Poland \label{UWarsaw} \and
Aix Marseille Universit\'e, CNRS/IN2P3, CPPM, Marseille, France \label{CPPM} \and
Instytut Fizyki J\c{a}drowej PAN, ul. Radzikowskiego 152, 31-342 Krak{\'o}w, Poland \label{IFJPAN} \and
Funded by EU FP7 Marie Curie, grant agreement No. PIEF-GA-2012-332350 \label{CurieChaves}  \and
School of Physics, University of the Witwatersrand, 1 Jan Smuts Avenue, Braamfontein, Johannesburg, 2050 South Africa \label{WITS} \and
Laboratoire d'Annecy de Physique des Particules, Universit\'{e} Savoie Mont-Blanc, CNRS/IN2P3, F-74941 Annecy-le-Vieux, France \label{LAPP} \and
Landessternwarte, Universit\"at Heidelberg, K\"onigstuhl, D 69117 Heidelberg, Germany \label{LSW} \and
Universit\'e Bordeaux, CNRS/IN2P3, Centre d'\'Etudes Nucl\'eaires de Bordeaux Gradignan, 33175 Gradignan, France \label{CENB} \and
Oskar Klein Centre, Department of Physics, Stockholm University, Albanova University Center, SE-10691 Stockholm, Sweden \label{OKC} \and
Wallenberg Academy Fellow \label{FellowConrad}  \and
Institut f\"ur Astronomie und Astrophysik, Universit\"at T\"ubingen, Sand 1, D 72076 T\"ubingen, Germany \label{IAAT} \and
Laboratoire Leprince-Ringuet, Ecole Polytechnique, CNRS/IN2P3, F-91128 Palaiseau, France \label{LLR} \and
APC, AstroParticule et Cosmologie, Universit\'{e} Paris Diderot, CNRS/IN2P3, CEA/Irfu, Observatoire de Paris, Sorbonne Paris Cit\'{e}, 10, rue Alice Domon et L\'{e}onie Duquet, 75205 Paris Cedex 13, France \label{APC} \and
Univ. Grenoble Alpes, CNRS, IPAG, F-38000 Grenoble, France \label{Grenoble} \and
Department of Physics and Astronomy, The University of Leicester, University Road, Leicester, LE1 7RH, United Kingdom \label{Leicester} \and
Nicolaus Copernicus Astronomical Center, Polish Academy of Sciences, ul. Bartycka 18, 00-716 Warsaw, Poland \label{NCAC} \and
Institut f\"ur Physik und Astronomie, Universit\"at Potsdam,  Karl-Liebknecht-Strasse 24/25, D 14476 Potsdam, Germany \label{UP} \and
Friedrich-Alexander-Universit\"at Erlangen-N\"urnberg, Erlangen Centre for Astroparticle Physics, Erwin-Rommel-Str. 1, D 91058 Erlangen, Germany \label{ECAP} \and
DESY, D-15738 Zeuthen, Germany \label{DESY} \and
Obserwatorium Astronomiczne, Uniwersytet Jagiello{\'n}ski, ul. Orla 171, 30-244 Krak{\'o}w, Poland \label{UJK} \and
Centre for Astronomy, Faculty of Physics, Astronomy and Informatics, Nicolaus Copernicus University,  Grudziadzka 5, 87-100 Torun, Poland \label{NCUT} \and
Department of Physics, University of the Free State,  PO Box 339, Bloemfontein 9300, South Africa \label{UFS} \and
Heisenberg Fellow (DFG), ITA Universit\"at Heidelberg, Germany \label{FellowRieger} \and
Department of Physics, Rikkyo University, 3-34-1 Nishi-Ikebukuro, Toshima-ku, Tokyo 171-8501, Japan \label{Rikkyo} \and
Japan Aerpspace Exploration Agency (JAXA), Institute of Space and Astronautical Science (ISAS), 3-1-1 Yoshinodai, Chuo-ku, Sagamihara, Kanagawa 229-8510,  Japan \label{JAXA} \and
Now at The School of Physics, The University of New South Wales, Sydney, 2052, Australia \label{MaxtedNowAt} \and
Now at Instituto de F\'{i}sica de S\~{a}o Carlos, Universidade de S\~{a}o Paulo, Av. Trabalhador S\~{a}o-carlense, 400 - CEP 13566-590, S\~{a}o Carlos, SP, Brazil \label{VianaNowAt} \and
Now at Kavli Institute for Particle Astrophysics and Cosmology, Department of Physics and SLAC National Accelerator Laboratory, Stanford University, Stanford, California 94305, USA \label{MeyerNowAt} \and
Department of Physics, Nagoya University, Furo-cho, Chikusa-ku, Nagoya 464-8601, Japan \label{Nagoya} \and
Nobeyama Radio Observatory, Minamimaki, Minamisaku, Nagano 384-1805 \label{Nobeya}
}

\date{Received 08 February 2017 / Accepted 24 October 2017}

\abstract{The H.E.S.S. collaboration has discovered a new very high energy (VHE, E $>$ 0.1 TeV) $\gamma$-ray source, HESS\,J1741$-$302, located in the Galactic plane. 
Despite several attempts to constrain its nature, no plausible counterpart has been found so far at X-ray and MeV/GeV $\gamma$-ray energies, and the source remains 
unidentified. An analysis of 145-hour of observations of HESS\,J1741$-$302 at VHEs has revealed a steady and relatively weak TeV source ($\sim$1$\%$ of the Crab Nebula 
flux), with a spectral index of $\Gamma$ = 2.3 $\pm$ 0.2$_{\text{stat}}$ $\pm$ 0.2$_{\text{sys}}$, extending to energies up to 10 TeV without any clear signature of a 
cut-off. In a hadronic scenario, such a spectrum implies an object with particle acceleration up to energies of several hundred TeV. Contrary to most H.E.S.S. 
unidentified sources, the angular size of HESS\,J1741$-$302 is compatible with the H.E.S.S. point spread function at VHEs, with an extension constrained to be below 
0.068$^{\circ}$ at a 99$\%$ confidence level. The $\gamma$-ray emission detected by H.E.S.S. can be explained both within a hadronic scenario, due to collisions 
of protons with energies of hundreds of TeV with dense molecular clouds, and in a leptonic scenario, as a relic pulsar wind nebula, possibly powered by the middle-aged 
(20 kyr) pulsar PSR\,B1737$-$30. A binary scenario, related to the compact radio source 1LC\,358.266$+$0.038 found to be spatially coincident with the best fit position 
of HESS\,J1741$-$302, is also envisaged.} 


\offprints{H.E.S.S.~collaboration, \\
\email{\href{contact.hess@hess-experiment.eu}{contact.hess@hess-experiment.eu}} \\
\protect\\\protect\footnotemark[1] Corresponding authors
\protect\\\protect\footnotemark[2] Deceased
}

\maketitle

\makeatletter
\renewcommand*{\@fnsymbol}[1]{\ifcase#1\@arabic{#1}\fi}
\makeatother
 
\section{Introduction} \label{intro}

The H.E.S.S. Galactic Plane Survey (HGPS) catalog \citep{2017a} contains 78 sources of VHE $\gamma$-rays, of which nearly 90$\%$ have been associated with at least one 
plausible counterpart in multi-wavelength catalogs. Pulsar wind nebulae (PWNe) appear to be the most prevalent VHE source class in the Galaxy, but there are also a sizable 
number of associations with supernova remnants (SNRs) and a handful of $\gamma$-ray binary systems. Many of the VHE sources have multiple associations, however, and it 
remains a major challenge to disentangle these and pinpoint the physical origin of the observed emission. Currently less than half of the sources in the HGPS catalog can 
be considered firmly identified. HESS J1741$-$302 is one of the most challenging VHE sources in this regard.

A preliminary detection of HESS\,J1741$-$302 was previously announced by H.E.S.S. \citep{tibolla2008} with an integrated flux level of $\sim$1$\%$ of the Crab Nebula flux 
above 1 TeV, HESS\,J1741$-$302 is not only unidentified but also one of the $\sim$10$\%$ of VHE sources lacking even a promising association in the HGPS catalog. Thanks 
to the increased amount of high-quality VHE data and improved analysis techniques employed in this study, leading to a better sensitivity and angular resolution with respect 
to the previous analysis, the morphology and spectrum of HESS\,J1741$-$302 can now be characterised in detail for the first time, and opens up the possibility for deeper 
multi-wavelength follow-up observations of the source.

The region around HESS\,J1741$-$302 is rather complex as shown in Fig. \ref{excess}, harbouring two pulsars\footnote{Note that there is another pulsar, PSR\,J1741$-$3016, 
located $\sim$0.1$^{\circ}$ away from the best fit position of HESS\,J1741$-$302. The VHE emission scenarios related to this pulsar are excluded since it is extremely old 
($\tau_{\text{c}}$\,=\,$\sim$3.3 Myr) and has a very low spin-down luminosity of 5.2 $\times$ $10^{31}$ erg/s.} as listed in Table \ref{pulsarTable}, the OH/IR star 
OH\,358.23$+$0.11 \citep{ohir}, a binary system, WR98a \citep{monnier}, which is included in the Particle-Accelerating Colliding Wind Binaries catalog \citep{pacwb}, and a 
luminous blue variable (LBV) star\footnote{Note that there are less than 20 LBV stars known in the Galaxy with their remarkable mass-loss rates, while there are only a few 
TeV sources found in the vicinity of LBV stars \citep{gavin}.}, Wray\,17$-$96 \citep{lbv}. In addition, a compact radio source, 1LC\,358.266$+$0.038 \citep{330MHZ} and a 
variable star, 2MASS J17411910$-$3023043\footnote{Note that the the 2MASS star and the compact radio object are not positionally coincident.} \citep{2mass}, are found to be 
spatially coincident with the best fit position of HESS\,J1741$-$302. This non-thermal radio source, whose nature is still uncertain, is most likely a synchrotron emitter, 
displaying a spectral index of $\alpha$ = 1.1 in its integrated flux density (when S\,$\propto\nu^{-\alpha}$). No variability is known for the compact radio object 
1LC\,358.266$+$0.038. On the other hand, the spectral energy distribution of the variable 2MASS star indicates that it might be a late type variable star.

\begin{table*} 
\renewcommand{\arraystretch}{1.3}
\caption{Pulsars around HESS\,J1741$-$302 with their properties. Pulsar data were taken from \cite{ATNF}. The position column indicates the positions of pulsars, while the 
columns labelled $\tau_{c}$ and $D_{\text{PSR}}$ show the characteristic age and the pulsar dispersion measure distance, respectively. The offset and separation columns 
show the relative offset of the pulsars with respect to the best fit position of HESS\,J1741$-$302 and the corresponding physical separation for the given pulsar distances, 
respectively. The $\dot{E}$ and $\dot{E}/D_{\text{PSR}}^{2}$ columns give the spin-down luminosity and flux of pulsars, respectively. The column labelled as 
$L_{\gamma}/\dot{E}$ gives the VHE $\gamma$-ray efficiencies of pulsars.}  
\label{pulsarTable}
\centering 
\begin{tabular}{c c c c c c c c c} 
\hline\hline 
Pulsar Name & Position & $\tau_{\text{c}}$ & $D_{\text{PSR}}$ & Offset        & Separation & $\dot{\text{E}}$ & $\dot{E}$/$D_{\text{PSR}}^{2}$ & $L_{\gamma}/\dot{\text{E}}$ \\ 
            & (J2000)  & (kyr)             &   (kpc)          & ($^{\circ}$)  & (pc)       & (erg/s)          & (erg/s/kpc$^{2}$)              & ($\%$)                      \\
\hline 

PSR\,B1737$-$30      & R.A.: 17$^{\text{h}}$40$^{\text{m}}$33.8$^{\text{s}}$   & 20.6   & 0.4    & 0.19 & 1.3  & 8.2 $\times$ $10^{34}$ & 5.1 $\times$ $10^{35}$ & 0.03     \\
                    & Dec.: $-$30$^{\circ}$15$^{\prime}$43.2$^{\prime\prime}$ &        & 3.28   &      & 10.9 &                     & 7.6 $\times$ $10^{33}$ & 1.84      \\ \hline
PSR\,J1739$-$3023    & R.A.: 17$^{\text{h}}$39$^{\text{m}}$39.8$^{\text{s}}$   & 159    & 3.41   & 0.35 & 20.8 & 3.0 $\times$ $10^{35}$ & 2.6 $\times$ $10^{34}$ & 0.54     \\    
                    & Dec.: $-$30$^{\circ}$23$^{\prime}$12.0$^{\prime\prime}$ &        &        &      &      &                     &                     &           \\ 

\hline 

\end{tabular}
\end{table*}

Dedicated observations of HESS\,J1741$-$302 at X-rays and high-energy (HE, 0.1$-$100 GeV) $\gamma$-rays did not result in any obvious counterpart. The source region has 
been observed by the {\it Suzaku} \citep{suzakuTel}, {\it Chandra} \citep{chandraTel} and {\it Swift} X-ray \citep{swiftTel} telescopes. The {\it Suzaku} observations 
\citep{suzaku} showed no significant X-ray emission neither from the HESS\,J1741$-$302 region nor from the position of PSR\,B1737$-$30 after investigation of 46.2 ks of 
data. In that study, the X-ray flux upper limits for the HESS\,J1741$-$302 region and the pulsar were provided as $F_{\text{X-ray, HESS\,J1741$-$302}}$ ($1-9$ keV) $<$ 
1.6 $\times$ $10^{-13}$ erg s$^{-1}$ cm$^{-2}$ and $F_{\text{X-ray, PSR\,B1737$-$30}}$ ($1-9$ keV) $<$ 3.5 $\times$ $10^{-13}$ erg s$^{-1}$ cm$^{-2}$ at a 90$\%$ confidence 
level, respectively. A recent investigation of the {\it Chandra} X-ray data\footnote{Two different dataset of 19.7 ks and 44.4 ks, coming from different observation regions, 
were investigated.} \citep{mwl1741} did not yield any obvious counterpart for the TeV emission. The X-ray source {\it CXOU} J174115.2$-$302434, which is confidently 
classified as a star, is located within 2$\sigma$ error interval of the best fit position of HESS\,J1741$-$302. In that study, scenarios related to active galactic nuclei 
were ruled out due to low X-ray to TeV flux ratio.

At HE $\gamma$-rays, there is no catalogued source coincident with the position of the H.E.S.S. source \citep{fermi3FGL, fermi2FHL}. A recent investigation of the 
{\it Fermi} LAT data has revealed a new HE source Fermi\,J1740.1$-$3013 \citep{fermiJ1741} which is $\sim$0.3$^{\circ}$ offset from the best fit position of 
HESS\,J1741$-$302. In that study, the authors suggested a possible association between the HE source and both the pulsars PSR\,B1737$-$30 and PSR\,J1739$-$3023 since 
the curved spectrum of the new HE source resembles $\gamma$-ray pulsars, although the spectral parameters could not be tightly constrained.

In this paper, the analysis of a large 145-hour VHE dataset on HESS\,J1741$-$302, obtained in part due to a relative proximity to the deep observed Galactic Center, are 
presented with the aim to constrain the nature of this ``hidden accelerator''. Section \ref{hess} describes the details of VHE data analysis and presents the results. Section 
\ref{mcs} describes a dedicated analysis of $^{12}$CO and HI data from the interstellar medium along the line of sight toward HESS\,J1741$-$302, which has been performed to 
investigate the possible hadronic origin of the observed emission. The results are discussed in Section \ref{interpret}, in which the potential origin of the faint VHE emission 
from HESS\,J1741$-$302 is considered in detail, with particular focus on the underlying particle acceleration mechanisms responsible for the production of $\gamma$-rays in both 
hadronic and leptonic scenarios. 

\section{H.E.S.S. Data Analysis and Results} \label{hess}

\subsection{The H.E.S.S. Telescopes}

The High Energy Stereoscopic System (H.E.S.S.) is an array of five imaging atmospheric Cherenkov telescopes located in the Khomas Highland of Namibia, 1800 m above sea 
level. H.E.S.S. in phase I comprised four 12 m diameter telescopes which have been fully operational since 2004. A fifth telescope (CT5), with a larger mirror diameter of 
28 m and newly designed camera \citep{hess2}, was added in the center of the array and has been operational since September 2012. The H.E.S.S. phase I array configuration 
is sensitive to $\gamma$-ray energies between 100 GeV and several tens of TeV. The VHE H.E.S.S. data presented in this paper were taken with the H.E.S.S. phase I array 
configuration, which can measure extensive air showers with an angular resolution better than ${0.1}^{\circ}$ and an average energy resolution of 15$\%$ for an energy of 
1 TeV \citep{Aharonian06}. 

\subsection{Detection and Morphological Analysis} \label{morph}

\begin{figure*}
\centering
\includegraphics[width=17cm]{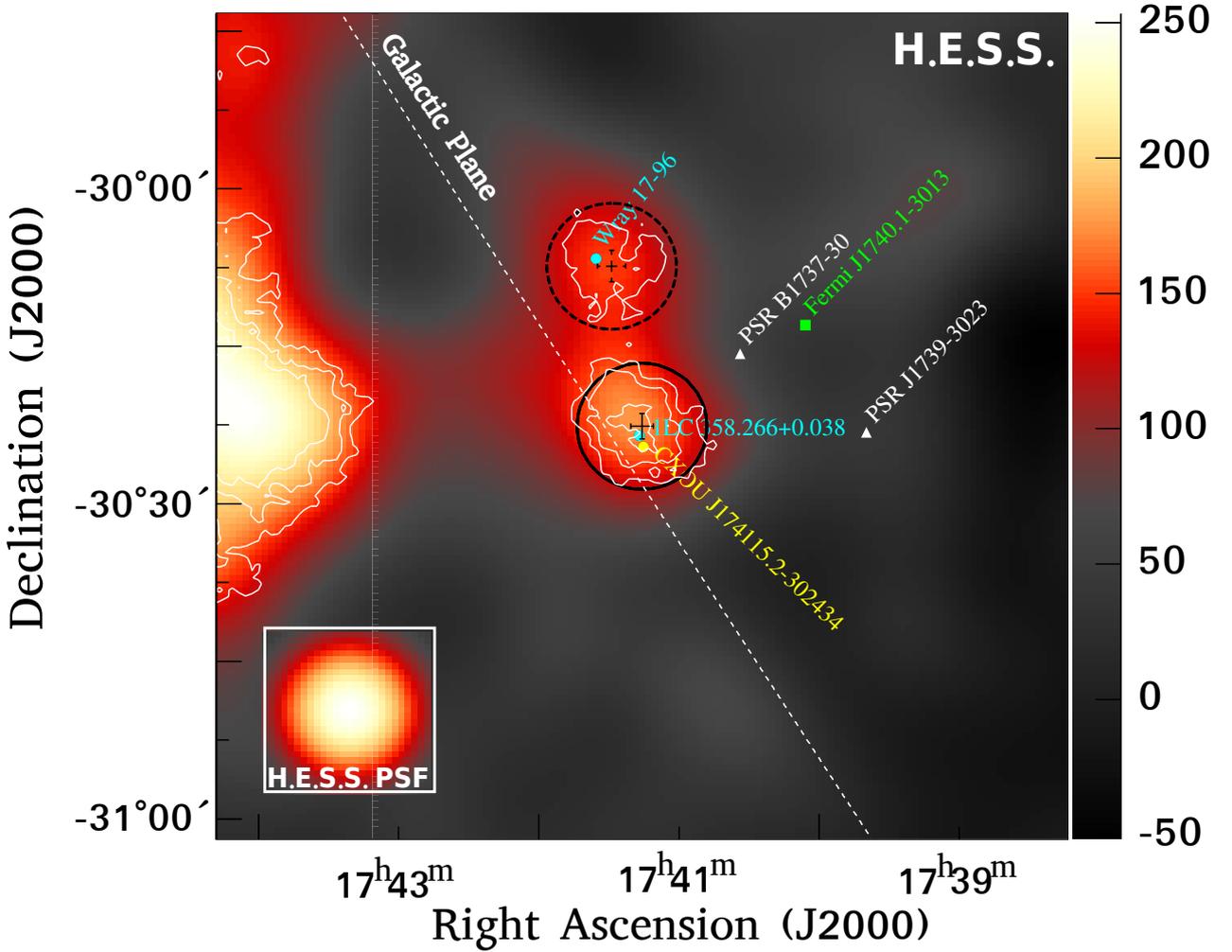}
\caption{Shown in this figure is the $\gamma$-ray excess map of the region around HESS\,J1741$-$302, smoothed with a Gaussian kernel of 0.08$^{\circ}$ width chosen to match 
the H.E.S.S. PSF which is shown inside the white box in the bottom left corner. The color scale is in units of excess $\gamma$-rays per smoothing Gaussian width. The black 
circle shows the source region used for extracting the spectrum, while the black cross indicates the value and 1$\sigma$ error interval (including systematic uncertainties) 
of the best fit position of the source. Similarly, the dashed black circle shows the hotspot region, while the dashed black cross indicates the value and 1$\sigma$ error 
interval of the best fit position of this emission. The pulsars listed in Table \ref{pulsarTable} are marked with white triangles. The light blue dots mark the position of 
the compact radio source 1LC\,358.266$+$0.038 and the LBV star Wray\,17$-$96, while the green dot marks the best fit position of Fermi\,J1740.1$-$3013, respectively. 
The yellow circle show the location of the Chandra X-ray source {\it CXOU} J174115.2$-$302434. The white dashed line indicates the orientation of the Galactic plane, while 
the white contours show 5$\sigma$, 6$\sigma$ and 7$\sigma$ (pre-trial) statistical significance of the VHE $\gamma$-ray emission.}
\label{excess}
\end{figure*}

The observations of the field of view around HESS\,J1741$-$302 were carried out between 2004 and 2013, corresponding to an acceptance corrected live-time of 145 h of 
H.E.S.S. phase I data after the application of the quality selection criteria \citep{Aharonian06}. The data have been analyzed with the H.E.S.S. analysis package for shower 
reconstruction and the multivariate analysis technique \citep{TMVA} has been applied for providing an improved discrimination between hadrons and $\gamma$-rays. Using this 
dataset, HESS\,J1741$-$302 is detected with a statistical significance of 7.8$\sigma$ pre-trials \citep{lima}, which corresponds to a 5.4$\sigma$ post-trial. In order to 
provide improved angular resolution and reduce contamination from the bright and nearby source HESS\,J1745$-$303 \citep{scanA}, the source position and morphology have been 
obtained with a hard cut configuration which requires a minimum of 160 photo-electrons per image. The cosmic-ray background level was estimated using the ring background 
model \citep{berge2007}. Figure \ref{excess} shows the acceptance corrected and smoothed VHE $\gamma$-ray excess map of the region around HESS\,J1741$-$302, also indicating 
the locations of known astronomical objects.

\begin{table*}  
\renewcommand{\arraystretch}{1.3}
\caption{Results of the HESS\,J1741$-$302 morphology analysis. The column labelled Morphology Model shows the model components used for modelling the HESS\,J1741$-$302 
region, while the columns labelled R.A., Dec. and Extension indicate the best fit position and the extension of the corresponding model component, respectively. The 
LogLikelihood column gives the log-likelihood value of the combined model used for comparing the tested models, while the $\chi^{2}$/ndf column gives the chi-square test 
for goodness of fit along with the corresponding p-value. Note that each model component has four free parameters.} 
\label{morphtable}
\centering 
\begin{tabular}{c c c c c c} 
\hline
\hline 

Morphology Model       & R.A.    & Dec.     & Extension     & LogLikelihood  & $\chi^{2}$/ndf \\ 
                       & (J2000) & (J2000)  & ($^{\circ})$  &                & (p-value) \\

\hline 

\underline{Model A}    &                                                                                       &                                                                                            &                                   & 23677.2 & 14783.3/14653  \\
HESS\,J1741$-$302      & 17$^{\text{h}}$41$^{\text{m}}$15.4$^{\text{s}}$ $\pm$ 3.6$^{\text{s}}_{\text{stat}}$  & $-$30$^{\circ}$22$^{\prime}$37.4$^{\prime\prime}$ $\pm$ 51$^{\prime\prime}_{\text{stat}}$  & 0.032 $\pm$ 0.018$_{\text{stat}}$ &         & (0.22) \\ 
Contamination          & 17$^{\text{h}}$42$^{\text{m}}$49.8$^{\text{s}}$ $\pm$ 42.2$^{\text{s}}_{\text{stat}}$ & $-$30$^{\circ}$09$^{\prime}$56.4$^{\prime\prime}$ $\pm$ 4.4$^{\prime}_{\text{stat}}$       & 0.409 $\pm$ 0.089$_{\text{stat}}$ &         &  \\ 

\hline

\underline{Model B}    &                                                                                       &                                                                                            &                                   & 23664.9 & 14764.4/14649  \\
HESS\,J1741$-$302      & 17$^{\text{h}}$41$^{\text{m}}$15.8$^{\text{s}}$ $\pm$ 3.6$^{\text{s}}_{\text{stat}}$  & $-$30$^{\circ}$22$^{\prime}$37.2$^{\prime\prime}$ $\pm$ 50$^{\prime\prime}_{\text{stat}}$  & 0.041 $\pm$ 0.018$_{\text{stat}}$ &         & (0.25) \\
Contamination          & 17$^{\text{h}}$43$^{\text{m}}$46.3$^{\text{s}}$ $\pm$ 79.2$^{\text{s}}_{\text{stat}}$ & $-$30$^{\circ}$07$^{\prime}$43.1$^{\prime\prime}$ $\pm$ 6.8$^{\prime}_{\text{stat}}$       & 0.498 $\pm$ 0.121$_{\text{stat}}$ &     & \\
Hotspot structure      & 17$^{\text{h}}$41$^{\text{m}}$28.8$^{\text{s}}$ $\pm$ 4.5$^{\text{s}}_{\text{stat}}$  & $-$30$^{\circ}$07$^{\prime}$22.5$^{\prime\prime}$ $\pm$ 66$^{\prime\prime}_{\text{stat}}$  & 0.019 $\pm$ 0.012$_{\text{stat}}$ &     & \\ 

\hline 

\end{tabular}
\end{table*}

For the morphology analysis, various models, including symmetric and asymmetric 2D Gaussian functions and a combination of these, convolved with the H.E.S.S. point spread 
function (PSF), were fitted to the excess map by using the Sherpa fitting package \citep{sherpa}. The H.E.S.S. PSF used in the morphology analysis was modelled as a weighted 
sum of three 2D Gaussian functions. The morphology models were compared using a log-likelihood ratio test (LLRT). The results of the morphology analysis are given in Table 
\ref{morphtable} along with the tested morphology models. The tested morphology models include three 2D Gaussians: one for describing the contamination coming from 
HESS\,J1745$-$303 (tagged as ``Contamination'' in Table \ref{morphtable}), one for HESS\,J1741$-$302 and the other one for compensating the hotspot structure seen in the 
analysis residual map after fitting the previous two components (Model A). Although Model B improved the fit by 4.0$\sigma$ (from LLRT) with respect to Model A, 
this improvement is not enough to claim the hotspot structure\footnote{The hotspot structure corresponds to the HESS\,J1741$-$302 A region mentioned in e.g. \cite{mwl1741} 
and \cite{fermiJ1741}.} as a new VHE $\gamma$-ray source, which would require a detection at a statistical significance above 5$\sigma$ after trial corrections. This 
hotspot feature is interesting but after trials consistent with a statistical fluctuation, thus requiring deeper observations to confirm. The best fit centroid position of 
the 2D Gaussian representing HESS\,J1741$-$302 is R.A.\@ (J2000):\@ 17$^{\text{h}}$41$^{\text{m}}$15.4$^{\text{s}}$ $\pm$ 3.6$^{\text{s}}_{\text{stat}}$ $\pm$ 
1.3$^{\text{s}}_{\text{sys}}$ and Dec.\@ (J2000):\@ $-$30$^{\circ}$22$^{\prime}$37.4$^{\prime\prime}$ $\pm$ 51$^{\prime\prime}_{\text{stat}}$ $\pm$ 
20$^{\prime\prime}_{\text{sys}}$. No significantly extended emission could be detected and HESS\,J1741$-$302 is found to be a point source with an extension upper limit of 
0.068$^{\circ}$ at a 99$\%$ confidence level.    

\subsection{Spectral Analysis} \label{spect}

\begin{figure}
\resizebox{\hsize}{!}{\includegraphics{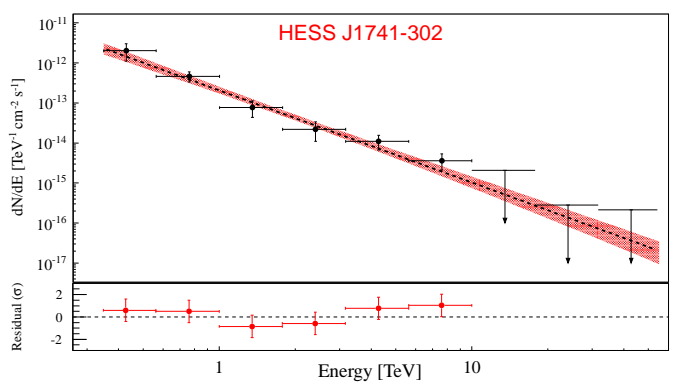}}
\caption{VHE $\gamma$-ray spectrum of HESS\,J1741$-$302 extracted from the source region shown in Fig. \ref{excess}. The black dots show the flux points with 1$\sigma$ 
errors, while the red shaded region represents the 68$\%$ confidence interval for the fitted spectral model (between 0.4 TeV and 56.2 TeV). The dashed black line shows the 
best fit power-law function. The differential flux upper limits are at the 99$\%$ confidence level and are shown with arrows. The spectrum shown is binned in such a way 
that all points have a minimum significance level of 2.0$\sigma$. The significance of the last point is 2.8$\sigma$.}
\label{spectrum}
\end{figure} 

A circular region with a radius of 0.1$^{\circ}$ centered at the best fit position of HESS\,J1741$-$302 (shown in Fig. \ref{excess}) was used as an integration region for 
extracting the VHE $\gamma$-ray spectrum of the source, and the reflected background model \citep{berge2007} was used for the background estimation. The differential VHE 
$\gamma$-ray spectrum was derived using the forward folding technique \citep{piron2001}, and is well described (p-value of 0.71) by a simple power-law function $dN/dE$ = 
$\Phi_{0}$ ($E/1\text{ TeV}$)$^{-\Gamma}$ as shown in Fig. \ref{spectrum}. The photon index of HESS\,J1741$-$302 is $\Gamma$ = 2.3 $\pm$ 0.2$_{\text{stat}}$ $\pm$ 
0.2$_{\text{sys}}$, while the normalization at 1 TeV is $\Phi_{0}$ = (2.1 $\pm$ 0.4$_{\text{stat}}$ $\pm$ 0.4$_{\text{sys}}$) $\times$ $10^{-13}$ cm$^{-2}$ s$^{-1}$ 
TeV$^{-1}$. The integral flux level above 1 TeV is $\Phi$($>$ 1\,TeV) = (1.7 $\pm$ 0.3$_{\text{stat}}$ $\pm$ 0.3$_{\text{sys}}$) $\times$ $10^{-13}$ cm$^{-2}$ s$^{-1}$ and 
corresponding to $\sim$1$\%$ of the Crab Nebula flux. The integrated energy flux $F_{\gamma}(> 0.4 \text{ TeV})$ is $\sim$1.2 $\times$ $10^{-12}$ erg cm$^{-2}$ s$^{-1}$. A 
power-law function with an exponential cut-off does not statistically improve the fit (0.4$\sigma$ from LLRT with respect to the power-law function). However, one can not 
exclude a cut-off in the $\gamma$-ray spectrum because of the limited statistics above 10 TeV.
 
Since this source does not show any significant extension, a variability analysis was performed to test its potential as a binary. The integral flux is found to be constant 
(p-value of 0.74 for the monthly light curve) within the H.E.S.S. dataset. Variability could be seen neither from run-wise ($\sim$30 min., p-value of 0.99) nor from 
year-wise (p-value of 0.33) light curves.

\section{Study of the Interstellar Medium} \label{mcs}

Dense gas regions can provide rich target material for accelerated particles to produce VHE $\gamma$-ray emission \citep{aha90}. For investigating a possible hadronic 
origin of the emission detected from the direction of HESS\,J1741$-$302, the analysis of data from available surveys of atomic and molecular hydrogen around the location of 
the source has been carried out.  

\begin{figure} 
\resizebox{\hsize}{!}{\includegraphics{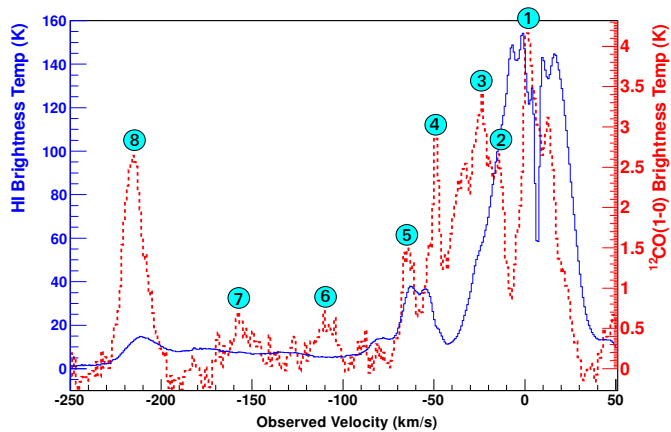}}
\caption{Brightness temperature distribution of molecular and atomic gas along the line-of-sight to HESS\,J1741$-$302. The red dashed line indicates $^{12}\text{CO}$ data 
from the Nanten survey, while the blue line shows HI data from the SGPS. The profiles are extracted from the circular region with a radius of 0.1$^{\circ}$ centered on the 
HESS\,J1741$-$302 best fit position. The abscissa shows the observed velocity of gas in km/s, while the ordinate shows the brightness temperature (the left axis for HI and 
the right axis for $^{12}\text{CO}$) in K. Peaks in the $^{12}\text{CO}$ brightness distribution are tagged with a number which indicates the corresponding interstellar 
cloud feature.}
\label{mcdata} 
\end{figure}

\begin{table*} 
\renewcommand{\arraystretch}{1.3}
\caption{Properties of interstellar cloud features along the line of sight to HESS\,J1741$-$302. The cloud number column shows the corresponding clouds from Fig. 
\ref{mcdata}, it also indicates whether the cloud is located at a near (N) or far (F) distance. The column labelled $V_{Int}$ shows the velocity integration intervals used 
to determine the mass of the cloud, while the columns labelled $V_{p}$ and $D$ give the velocity at which the maximum CO intensity occurs and the distance of the cloud from 
the Sun, respectively. The $M_{\text{cl}}$ column gives the mass of the cloud in terms of solar mass (M$_{\bigodot}$), while the column labelled $n_{Gas}$ gives the gas 
number density of the cloud. The mass uncertainties are around 50$\%$. The $M_{5}$/$D^{2}$ column gives the $\gamma$-ray visibility parameter of the cloud where $M_{5}$ = 
$M_{\text{cl}}$/$10^{5}$M$_{\bigodot}$ is the diffuse mass of the cloud located at a distance $D_{\text{kpc}}$ = $D$/1 kpc. The column labelled $k_{CR}$ gives the calculated 
CR density factor of a cloud, while the $W_{pp}$ column gives the total energy required in protons. The table is sorted according to the descending values of 
$M_{5}$/$D_{\text{kpc}}^{2}$.} 
\label{mctable} 
\centering 
\begin{tabular}{c c c c c c c c c} 
\hline
\hline 

Cloud Number & $V_{Int}$       & $V_{p}$  &    $D$    & $M_{\text{cl}}$       & $n_{Gas}$   & $M_{5}$/$D_{\text{kpc}}^{2}$  & $k_{CR}$ & $W_{pp}$\\ 
          &   (km/s)        & (km/s)   &   (kpc)   & ($M_{\bigodot}$)      & (cm$^{-3}$) &                  &          & (erg)\\

\hline 

3 (N)     & [$-$31, $-$18]  & $-$24.2  & 6.7       & 6.3 $\times$ $10^{4}$ & 380         & 0.0140                       & 53       & 1.5 $\times$ $10^{47}$ \\ 
1 (N)     & [$-$2, $+$7]    & $+$2.6   & 5.0       & 3.1 $\times$ $10^{4}$ & 460         & 0.0125                       & 60       & 7.0 $\times$ $10^{46}$ \\ 
(CMZ)     &                 &          & 8.8       & 9.7 $\times$ $10^{4}$ & 260         & 0.0125                       & 60       & 3.9 $\times$ $10^{47}$ \\
8 (N)     & [$-$221, $-$210]& $-$215.7 & 8.5       & 6.8 $\times$ $10^{4}$ & 200         & 0.0094                       & 79       & 4.7 $\times$ $10^{47}$\\
2 (N)     & [$-$19, $-$10]  & $-$15.7  & 6.0       & 2.8 $\times$ $10^{4}$ & 240         & 0.0078                       & 95       & 1.9 $\times$ $10^{47}$\\ 
  (F)     &                 &          & 11.2      & 9.8 $\times$ $10^{4}$ & 130         & 0.0078                       & 95       & 1.3 $\times$ $10^{48}$\\
4 (N)     & [$-$51, $-$47]  & $-$48.8  & 7.5       & 2.4 $\times$ $10^{4}$ & 100         & 0.0043                       & 174      & 7.3 $\times$ $10^{47}$\\ 
5 (N)     & [$-$68, $-$60]  & $-$64.2  & 7.7       & 2.3 $\times$ $10^{4}$ & 91          & 0.0039                       & 192      & 8.5 $\times$ $10^{47}$\\ 
6 (N)     & [$-$120, $-$100]& $-$109.6 & 8.3       & 1.9 $\times$ $10^{4}$ & 62          & 0.0028                       & 270      & 1.4 $\times$ $10^{48}$\\ 
7 (N)     & [$-$145, $-$165]& $-$154.6 & 8.4       & 2.0 $\times$ $10^{4}$ & 62          & 0.0028                       & 262      & 1.5 $\times$ $10^{48}$\\ 

\hline 

\end{tabular}
\end{table*}

The atomic hydrogen (HI) 21-cm (1420 MHz) spectral line data from the Southern Galactic Plane Survey (SGPS) \citep{sgps} and $^{12}$CO J=1$\rightarrow$0 rotational 
transition line emission data from the Nanten Galactic Plane Survey \citep{nanten} were investigated to determine the distribution of atomic and molecular hydrogen in the 
region of HESS\,J1741$-$302. The comparison of HI and $^{12}$CO brightness temperature along the line-of-sight is shown in Fig. \ref{mcdata}. The velocities of emission 
maxima (given in Table \ref{mctable}) were used to determine the distances to interstellar cloud features under the assumption of a Galactic rotation curve model 
\citep{grc}. These interstellar cloud features will be called ``clouds'' throughout the text. The near/far kinematic distance ambiguity (KDA) was resolved by using the HI 
self absorption and 21 cm continuum absorption as explained in e.g. \cite{kda}. Note that the non-zero kinematic velocity of Cloud 1 suggests that it is not far from the 
Central Molecular Zone (CMZ).

\begin{figure*} 

\centering
\includegraphics[width=17cm]{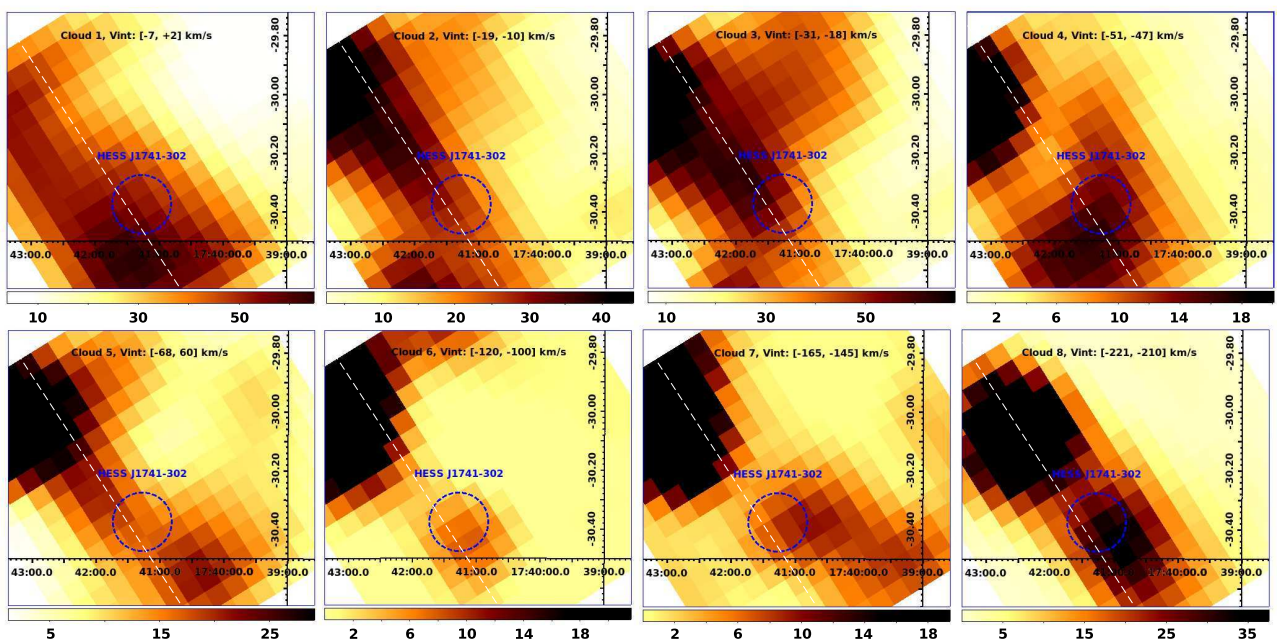}
\caption{Integrated $^{12}\text{CO}$ column density maps of the region around HESS\,J1741$-$302 for each cloud as labelled in Fig. \ref{mcdata}. Velocity integration intervals used 
for producing the corresponding column density map is given for each cloud. The dashed blue circles show the HESS\,J1741$-$302 region which is marked with a dashed black circle in 
Fig. \ref{excess}, while the white dashed line indicates the orientation of the Galactic plane. The color scale is in the units of $10^{20}$ cm$^{-2}$.}
\label{cdco1741} 
\end{figure*}

The column densities along the line-of-sight, averaged over the solid angle of the source region, were calculated by integrating the brightness temperature over the velocity intervals 
covering the observed peaks. The X-factors\footnote{The X-factor is described as the ratio of H$_{2}$ column density CO$^{12}$ intensity and the values used throughout this paper 
are average values in Galactic molecular clouds \citep{xfac}.}, $\text{X}_{\text{HI}}$ = 1.823 $\times$ $10^{18}$ cm$^{-2}$ (K km s$^{-1}$)$^{-1}$ and $\text{X}_{^{12}\text{CO}}$ = 1.5 $\times$ $10^{20}$ 
cm$^{-2}$ (K km s$^{-1}$)$^{-1}$, were taken from \cite{xhi} and \cite{xco}, respectively. Note that the HI X-factor assumes optically thin HI and may underestimate the HI column density by 
a factor 2$-$3 \citep{fukui}. The integrated $^{12}\text{CO}$ column density maps are shown in Fig. \ref{cdco1741} for each cloud. It was found out that the atomic-to-molecular gas ratio is 
$\sim$10$\%$, therefore the molecular gas component dominates. The estimated masses, distances and gas densities for each cloud along the line of sight to HESS\,J1741$-$302 are given in 
Table \ref{mctable} along with the calculated $M_{5}$/$D^{2}$ values for the cloud mass to distance squared, one of the principal parameters that determine the $\gamma$-ray visibility of a 
cloud, CR density factors\footnote{CR density factor is the ratio of CR density in a cloud to the one observed at the Earth. See \cite{aha90} for detailed discussions.} ($k_{CR}$) and total 
energy in protons ($W_{pp}$). Note that estimated mass values take into account only the $^{12}\text{CO}$ data. When taking into account the HI data, the mass values increase by a factor of 
10$\%$. As shown in Fig. \ref{cdco1741}, some of the clouds show a good correlation with the VHE emission observed from the direction of HESS\,J1741$-$302. The results obtained in 
this analysis will be used for discussing a possible hadronic scenario (see Sect. \ref{had}). 

\section{Discussion} \label{interpret}

HESS\,J1741$-$302 is one of the faintest VHE sources detected so far. Its energy spectrum extending up to at least 10 TeV suggests that the parental particle population producing such emission 
extends up to tens of TeV in the leptonic case or hundreds of TeV in the hadronic case.

Hereafter, various scenarios to explain the VHE emission from HESS\,J1741$-$302 are discussed: a hadronic scenario in which the emission arises from the interaction between an interstellar 
medium (ISM) gas target (like a cloud) and accelerated cosmic-rays, and a leptonic scenario, where VHE emission from a relic PWN by electron inverse Compton (IC) scattering off either cosmic 
microwave background (CMB) photons or infra-red (IR) photons emitted by the OH/IR star. A binary scenario related to the compact radio source 1LC 358.266$+$00.038, which is coincident with 
the best fit position of HESS\,J1741$-$302, can also be envisaged. 

\subsection{Hadronic Scenarios} \label{had}

The VHE spectrum of HESS\,J1741$-$302, extending up to 10 TeV, suggests that the VHE emission could be produced by protons accelerated up to hundreds of TeV colliding with the ambient dense 
gas. The total energy required in protons, $W_{pp}$, to produce the inferred $\gamma$-ray luminosity, $L_{\gamma}$ = 4$\pi F_{\gamma}(> 0.4\text{ TeV}) D^{2}$, can be estimated as 
$W_{pp} = L_{\gamma}$ $\times$ $t_{pp}$, where $t_{pp}$ = 5.76 $\times$ 10$^{15}$ $\times$ ($n_{gas}$/cm$^{-3}$)$^{-1}$ s is the cooling time for proton-proton collisions. Calculations of the 
energy budget suggest that $W_{pp}$ is between 7.0 $\times$ $10^{46}$ erg and 1.5 $\times$ $10^{48}$ erg depending on the density of the cloud of interest (see Table \ref{mctable} for 
$W_{pp}$ values and cloud distances, $D$, used for calculating $L_{\gamma}$). Concerning the energy budget, each interstellar cloud found along the line-of-sight to HESS\,J1741$-$302 could 
give rise to the observed VHE emission if an undetected CR source is located in the vicinity of (or inside) one of the clouds. In particular, Cloud 1, Cloud 3 and Cloud 8 in Fig. 
\ref{cdco1741} are promising ISM counterparts for the VHE emission as they partially overlap with HESS\,J1741$-$302. However, no established CR accelerator 
(such as a SNR) is close to, or coincident with, the source. 

The results presented in the study of diffuse $\gamma$-ray emission from the GC ridge \citep{Aharonian2006,arion} might suggest that the CRs accelerated by the galactic central source, with an age 
assumption of $\sim$10 kyr, can fill the region $|l|$ $<$ 1$^{\circ}$ but have not yet diffused beyond $|l|$ $\geq$ 1$^{\circ}$. The projected distance of the Cloud 8 (the closest one to the 
GC) is 260 pc $\pm$ 40 pc (corresponding to $l$ = $-1.7^{\circ}$). The scenario considering the illumination of Cloud 8 by CRs accelerated at the GC is excluded because of this large 
offset and since the required CR density factor in this cloud ($\sim$80) is significantly larger than that in the CMZ.  

\subsection{Leptonic Scenarios} \label{lep}

Electrons with energies of hundreds of TeV, undergoing IC scattering off CMB or ambient radiation fields, could potentially explain the emission from HESS\,J1741$-$302. 
These electrons could be accelerated at the wind termination shocks of the pulsars around HESS\,J1741$-$302 listed in Table \ref{pulsarTable}. Note that pulsars with 
$\tau_{c}$ $<$ 150 kyr and with $\dot{E}$/$D_{\text{PSR}}^{2}$ $\geq$ 10$^{34}$ erg s$^{-1}$ kpc$^{-2}$ are known to power PWNe that are detectable at very high energies 
\citep{pwnpopB}. Below, relic PWN scenarios related to the pulsar PSR\,B1737$-$30 will be discussed. Note that this pulsar have two distance approximations of 0.4 kpc and 
3.28 kpc coming from the absorption of pulsar emission by Galactic HI gas \citep{04kpc} and the dispersion measure \citep{328kpc}, respectively. Leptonic VHE emission 
scenarios related to the other pulsar, PSR\,J1739$-$3023, is excluded due to the very large offset of 0.35$^{\circ}$ with respect to the best fit position of 
HESS\,J1741$-$302.

The angular offset of $\sim$0.19$^{\circ}$ between the best fit position of HESS\,J1741$-$302 and PSR\,B1737$-$30 can be explained with minimum initial pulsar kick 
velocities of 65 km/s and 540 km/s for the pulsar distances of 0.4 kpc and 3.28 kpc, respectively. However, no proper motion in right ascension or in declination is known 
for this pulsar \citep{ATNF}. 

A calculation of the characteristic IC cooling time scales suggests that $\sim$60 TeV electrons can still exist after a time period that corresponds to the characteristic 
age of the pulsar, while diffusion times of electrons are negligible when compared to this time scale. In such a leptonic scenario, the total energy required in electrons 
that gives rise to the inferred $\gamma$-ray luminosity can be estimated as $W_{e}$($D_{\text{PSR}}$ = 0.4 kpc) = $5.4$ $\times$ $10^{44}$ erg or $\sim$1$\%$, and 
$W_{e}$($D_{\text{PSR}}$ = 3.28 kpc) = $3.6$ $\times$ $10^{46}$ erg or $\sim$70$\%$ of the maximum energy available from the pulsar when taking into account its age.  

In the case of a relic PWN scenario, one expects a (still) bright offset PWN, while the X-ray PWN close to the pulsar is already dim. Although X-ray observations could not 
detect any X-ray PWN around PSR\,B1737$-$30, the phenomenological approach given by \cite{mattana} predicts a faint X-ray PWN around the pulsar with an X-ray flux 
level\footnote{This flux level is estimated by using $\dot{E}$ and it increases to $F$(2$-$10 keV) = 1.4 $\times$ $10^{-13}$ erg s$^{-1}$cm$^{-2}$ when $\tau_{\text{c}}$ is 
taken into account. These expectations are compatible with flux upper limits given in \cite{suzaku}.} of $F$(2$-$10 keV) = 5.1 $\times$ $10^{-14}$ erg s$^{-1}$cm$^{-2}$ for 
$D_{\text{PSR}}$ = 0.4 kpc. Therefore, further targeted X-ray observations of PSR\,B1737$-$30 could potentially test the electron accelerator hypothesis.  

The spatially coincident compact radio source 1LC\,358.266$+$0.038 is a promising object that can give hints about the nature of HESS\,J1741$-$302. Although the spectral 
index of this source ($\alpha$ = 1.1) is slightly steeper with respect to the indices (0.2 $\leq$ $\alpha$ $\leq$ 1.0) suggested by \cite{fender}, a $\gamma$-ray binary 
origin could be envisaged as it was done for the $\gamma$-ray binary LS 5039 \citep{lsx}, while \cite{pynzar} classified this object as extragalactic. In view of the 
point-like nature of HESS\,J1741$-$302 and given the low statistics, the fact that no variability has been observed can not be taken as evidence for disfavoring a binary 
origin, while scenarios related to the extragalactic origin of the source are highly unlikely when taking into account that the observed VHE emission is located in the 
Galactic plane, $\sim1.7^{\circ}$ away from the Galactic Center.

Other leptonic scenarios can be considered taking into account the stars in the vicinity of HESS\,J1741$-$302. In such scenarios, the strong IR photon field of 
OH\,358.23$+$0.11 can provide target photons for the production of IC scattered $\gamma$-rays in the presence of relativistic electron populations provided by 
PSR\,B1737$-$30 in a putative OH/IR star - pulsar system. Alternatively, VHE $\gamma$-ray emission scenarios related to the $\sim$1.5 year period binary star WR\,98a can be 
discussed. In both of these cases, one expects that the center-of-gravity of the VHE $\gamma$-ray excess must be coincident with the region where VHE electron acceleration 
takes place. Using this argument, VHE $\gamma$-ray emission scenarios related to WR\,98a and OH\,358.23$+$0.11 are disfavored in the light of significant offsets 
($\sim$0.16$^{\circ}$ and $\sim$0.08$^{\circ}$ respectively) between these stars and the best fit position of HESS\,J1741$-$302. The LBV star Wray 17$-$96, located at a 
distance of 4.5 kpc, has an offset of $\sim$0.27$^{\circ}$ from the best fit position of HESS\,J1741$-$302 and can be excluded from the scenarios related to the origin of 
the source. On the other hand, this extremely rare type of star is located $\sim$1.1$\sigma$ away from the best fit position of the hotspot structure. 

\section{Conclusions} \label{conc}

H.E.S.S. has discovered a new unidentified VHE $\gamma$-ray source HESS\,J1741$-$302 in the Galactic plane. The analysis of H.E.S.S. data has shown that the $\gamma$-ray 
spectrum of HESS\,J1741$-$302 extends at least up to 10 TeV. Hadronic particles accelerated up to hundreds of TeV by an undetected cosmic-ray source interacting with the 
detected clouds along the line-of-sight to HESS\,J1741$-$302 can result in the observed VHE $\gamma$-ray emission, in particular, Cloud 1, Cloud 3 and Cloud 8 are promising 
candidates (see Sect. \ref{mcs}). On the other hand, a relic PWN scenario related to the pulsar PSR\,B1737$-$30 located at 0.4 kpc can not be excluded but is disfavoured due to the point-like 
nature of the source and its offset with respect to the pulsar. Future X-ray observations of this pulsar can be used for testing such a relic PWN scenario. A binary 
scenario related to the compact radio source 1LC\,358.266$+$0.038 is also possible. Despite a thorough investigation of its multi-wavelength environment and consideration 
of a variety of plausible emission mechanisms, this enigmatic VHE source remains unidentified. The future Cherenkov Telescope Array with its much better angular resolution 
and sensitivity will be able to further characterize the VHE $\gamma$-ray emission from this region.

\begin{acknowledgements}

The support of the Namibian authorities and of the University of Namibia in facilitating the construction and operation of H.E.S.S. is gratefully acknowledged, as is the support by the German 
Ministry for Education and Research (BMBF), the Max Planck Society, the German Research Foundation (DFG), the Alexander von Humboldt Foundation, the Deutsche Forschungsgemeinschaft, the French 
Ministry for Research, the CNRS-IN2P3 and the Astroparticle Interdisciplinary Programme of the CNRS, the U.K. Science and Technology Facilities Council (STFC), the IPNP of the Charles 
University, the Czech Science Foundation, the Polish National Science Centre, the South African Department of Science and Technology and National Research Foundation, the University of 
Namibia, the National Commission on Research, Science \& Technology of Namibia (NCRST), the Innsbruck University, the Austrian Science Fund (FWF), and the Austrian Federal Ministry for 
Science, Research and Economy, the University of Adelaide and the Australian Research Council, the Japan Society for the Promotion of Science and by the University of Amsterdam. We appreciate 
the excellent work of the technical support staff in Berlin, Durham, Hamburg, Heidelberg, Palaiseau, Paris, Saclay, and in Namibia in the construction and operation of the equipment. This 
work benefited from services provided by the H.E.S.S. Virtual Organisation, supported by the national resource providers of the EGI Federation. This research has made use of software provided 
by the {\it Chandra} X-ray Center (CXC) in the application packages CIAO, ChIPS, and Sherpa. This research has made use of the SIMBAD database, operated at CDS, Strasbourg, France. This 
research has made use of the ATNF pulsar catalog database (http://www.atnf.csiro.au/research/pulsar/psrcat/). The NANTEN project is based on the mutual agreement between Nagoya University and 
the Carnegie Institution of Washington. Sabrina Casanova and Ekrem O\u{g}uzhan Ang\"uner acknowledge the support from the Polish National Science Center under the Opus Grant 
UMO-2014/13/B/ST9/00945. 

\end{acknowledgements}

\bibpunct{(}{)}{;}{a}{}{,}

\bibliographystyle{aa} 
\bibliography{1741_AA_Draft_3RD-RESUBMIT} 

\end{document}